\documentclass[preprint,aps]{revtex4-1}
\usepackage{lineno}
\usepackage{graphicx}

\usepackage{soul}
\usepackage{epstopdf}
\usepackage{bm}
\usepackage{upgreek}
\usepackage{amsmath}
\usepackage{lipsum}
\usepackage{color,xcolor}
\usepackage{amsfonts}
\DeclareUnicodeCharacter{2009}{\,}
\DeclareUnicodeCharacter{2212}{\ensuremath{-}}
\DeclareUnicodeCharacter{03B4}{$\delta$}
\begin{document}

\title{Unusual Dual Flat Bands and two-dimensional Dirac-node Arc State in Kagome Metal Ni$_3$In$_2$S$_2$}
\author{Bo Liang$^{1,2,\sharp}$, Yichen Liu$^{3,\sharp}$, Jie Pang$^{1,2,\sharp}$, Hanbin Deng$^{4,\sharp}$, Taimin Miao$^{1,2}$, Wenpei Zhu$^{1,2}$, Neng Cai$^{1,2}$, Tiantian Zhang$^{5}$, Jiayu Liu$^{6}$, Zhicheng Jiang$^{7}$, Zhanfeng Liu$^{7}$, Hongen Zhu$^{7}$, Yuliang Li$^{7}$, Tongrui Li$^{7}$, Mingkai Xu$^{1,2}$, Hao Chen$^{1,2}$, Xiaolin Ren$^{1,2}$, Chaohui Yin$^{1,2}$, Yingjie Shu$^{1,2}$, Yiwen Chen$^{1,2}$, Yu-Tian Zhang$^{5}$, Zhengtai Liu$^{8}$, Dawei Shen$^{7}$, Mao Ye$^{8}$, Fengfeng Zhang$^{9}$, Shenjin Zhang$^{9}$, Shengtao Cui$^{7}$, Zhe Sun$^{7}$, Koji Miyamoto$^{10}$, Taichi Okuda$^{10}$, Kenya Shimada$^{10}$, Lihong Yang$^{1}$, Jia-Xin Yin$^{4,11}$, Lin Zhao$^{1,2,12}$, Zuyan Xu$^{9}$, Haijun Zhang$^{3,13,14,*}$, Youguo Shi$^{1,2,12,*}$, X. J. Zhou$^{1,2,12,*}$ and Guodong Liu$^{1,2,12,*}$
}

\affiliation{
\\$^{1}$Beijing National Laboratory for Condensed Matter Physics, Institute of Physics, Chinese Academy of Sciences, Beijing 100190, China.
\\$^{2}$School of Physical Sciences, University of Chinese Academy of Sciences, Beijing 100049, China.
\\$^{3}$National Laboratory of Solid State Microstructures and School of Physics, Nanjing University, Nanjing 210093, China.
\\$^{4}$Department of Physics and Guangdong Basic Research Center of Excellence for Quantum Science, Southern University of Science and Technology, Shenzhen 518055, China.
\\$^{5}$Institute of Theoretical Physics, Chinese Academy of Sciences, Beijing 100190, China.
\\$^{6}$Shanghai Institute of Microsystem and Information Technology, Chinese Academy of Sciences, Shanghai 200050, China.
\\$^{7}$National Synchrotron Radiation Laboratory and School of Nuclear Science and Technology, University of Science and Technology of China, Hefei 230026, China.
\\$^{8}$Shanghai Synchrotron Radiation Facility, Shanghai Advanced Research Institute, Chinese Academy of Sciences, Shanghai 201210, China.
\\$^{9}$Technical Institute of Physics and Chemistry, Chinese Academy of Sciences, Beijing 100190, China.
\\$^{10}$Hiroshima Synchrotron Radiation Center, Hiroshima University, Higashi-Hiroshima, Japan.
\\$^{11}$Quantum Science Center of Guangdong-Hong Kong-Macao Greater Bay Area (Guangdong), Shenzhen, China.
\\$^{12}$Songshan Lake Materials Laboratory, Dongguan, Guangdong 523808, China.
\\$^{13}$Jiangsu Physical Science Research Center, Nanjing 210093, China.
\\$^{14}$Jiangsu Key Laboratory of Quantum Information Science and Technology, Nanjing University, China.
\\$^{\sharp}$These people contribute equally to the present work.
\\$^{*}$Corresponding authors: gdliu\_arpes@iphy.ac.cn, xjzhou@iphy.ac.cn, ygshi@iphy.ac.cn, zhanghj@nju.edu.cn
}

\date{\today}



\maketitle

\newpage

\noindent{\bf Kagome materials are at the frontier of condensed matter physics. An ideal kagome lattice features only one geometrically frustrated flat band spanning the entire momentum space and a single Dirac cone at the Brillouin-zone corners. However, for the first time, here we observe unusual flat-band and Dirac physics in the newly discovered “322” kagome material Ni$_3$In$_2$S$_2$ by combining high-resolution synchrotron- and laser-based angle-resolved photoemission spectroscopy with a micro-focused beam, scanning tunneling microscopy, and first-principles calculations. We resolve two distinct electronic flat-band states located in close proximity to the Fermi level: a robust Topological Surface Flat Band at $\sim$40\,meV below the Fermi level on the Sulfur-terminated surface, originating from weak topological insulator states, and a kagome lattice-derived flat band at $\sim$100\,meV binding energy with an ultranarrow bandwidth ($\sim$5\,meV). Instead of the single Dirac cone, the Indium-terminated surface hosts a rare two-dimensional Dirac-node arc state, where the gapless Dirac nodes extend along an open one-dimensional line crossing the Brillouin-zone boundary, exhibiting sharp linear dispersion, exceptionally high Fermi velocity, and pronounced circular dichroism. These findings establish Ni$_3$In$_2$S$_2$ as a unique topological kagome metal in which multiple flat-band states of different physical origin coexist with an unusual Dirac-node arc, opening an avenue for discovering flat-band--driven and topology-enabled quantum phenomena.}

~\\

Flat bands provide a powerful route to realizing strongly correlated and topology-enabled quantum phenomena: by substantially suppressing electronic kinetic energy, they amplify Coulomb interaction and potentially drive spontaneous symmetry breaking and topological phase transitions. This interaction-driven mechanism underlies a wide range of emergent phases---including superconductivity, unconventional magnetism, and correlation-driven topological states---across heavy-fermion materials\cite{Wirth2016}, twisted moir\'{e} systems\cite{Bistritzer2011,Nuckolls2024}, and kagome lattices\cite{Yin2022Kagomereview,ClaudiaKagomereview}. Among these platforms, kagome lattices hold a special position, offering a unique geometrical setting for flat-band formation and naturally intertwining electronic correlations with topological characteristics through their frustrated geometry. In such a lattice, interaction and destructive phase interference among Bloch wavefunctions lead to the formation of Dirac fermions, saddle points, and flat bands. Such characteristic electronic structures, boosted by intricate interplay among lattice symmetry, magnetism, electron correlations, and spin--orbit coupling (SOC), manifest a wealth of physical phenomena, including quantum spin liquids\cite{Balents2010}, flat-band magnetism\cite{Lin2018,Yin2019}, nontrivial band topology\cite{Sun2011,Liu2019,Kang2020b,Zhang2022}, density waves\cite{Chen2021,Teng2022}, unconventional superconductivity\cite{Ortiz2020}, fractional quantum Hall effect (FQHE)\cite{Tang2011}, and quantum anomalous Hall effect (QAHE)\cite{Xu2015}. 
 
Several kagome families---such as the shandite-type M$_3$A$_2$B$_2$ compounds\cite{Liu2019,Zhang2022} (M = Co, Ni; A = In, Sn; B = S, Se), the AT$_3$X$_5$ family\cite{Ortiz2019,Yang2023,Liu2024Superconductivity} (A = K, Rb, Cs; T = V, Ti, Cr; X = Sb, Bi), the RT$_6$X$_6$ family\cite{Yin2020,DiSante2023,Song2025} (R = rare-earth elements; T = V, Cr, Mn, Fe, Co; X = Ge, Sn), and the pyrochlore metals\cite{Wakefield2023,HJW2024}---have revealed rich correlated and topological behaviour. However, direct spectroscopic identification of kagome lattice-derived flat bands coexisting with topological electronic states near the Fermi level ($E_F$) remains rare.

The “322” kagome family M$_3$A$_2$B$_2$ provides an ideal platform for addressing this challenge. In these materials, the 3d-orbital bands derived from the transition-metal atoms forming the kagome lattice intersect near the Fermi level, giving rise to topological nodal-line states in the absence of SOC. When the symmetry protection is broken or SOC is introduced, these nodal lines evolve into a variety of distinct topological quantum phases. A representative example is the ferromagnetic Weyl-kagome metal Co$_3$Sn$_2$S$_2$, where time--reversal symmetry breaking together with strong SOC gaps a mirror-protected nodal line and produces a pair of Weyl nodes with opposite chirality accompanied by surface Fermi arcs\cite{Liu2019}. Despite intensive study, a definitive spectroscopic observation of kagome-derived electronic states in Co$_3$Sn$_2$S$_2$ remains lacking, demonstrating the difficulty of accessing kagome-lattice features in the family of “322” kagome materials.

Ni$_3$In$_2$S$_2$, an isostructural analogue of Co$_3$Sn$_2$S$_2$, has recently attracted considerable interest due to theoretical predictions of a rare endless Dirac nodal-line state\cite{Zhang2022} and the observation of record-high carrier mobility among all known kagome materials, along with extremely large magnetoresistance (MR) in transport studies\cite{Fang2023,Zhang2024,KimPRB}. Remarkably, recent electrical transport and specific heat studies\cite{Das2024NIS} further point that, in addition to its high carrier mobility and extremely large magnetoresistance, Ni$_3$In$_2$S$_2$ may exhibit strong electronic correlations. Although two preliminary angle-resolved photoemission spectroscopy (ARPES) studies\cite{Zhang2022,Fang2023} have been conducted on Ni$_3$In$_2$S$_2$, the reported electronic properties remain highly controversial. In particular, the origin of the Dirac states at the $\bar{M}$ point and the topological nature of the material, including its proposed endless Dirac nodal-line character, are still under debate. Moreover, the lack of termination-resolved measurements has hindered a definitive assignment of its bulk and surface states, which is crucial in kagome materials where multiple cleaving surfaces naturally occur.

A comprehensive, high-resolution, and termination-sensitive investigation is therefore required to elucidate the intrinsic electronic structure of Ni$_3$In$_2$S$_2$, to determine whether kagome-derived electronic features---represented by flat bands---and topologically non-trivial electronic states indeed emerge near the Fermi level, and to clarify their respective microscopic origins within the broader “322” kagome family.

~\\
\noindent {\bf\large Results}\\
\noindent {\bf Crystal structure and bulk electronic properties of Ni$_3$In$_2$S$_2$.} The shandite-type compound Ni$_3$In$_2$S$_2$ crystallizes in space group No. 166 (R$\bar{3}$m). Each unit cell consists of three Ni-based kagome layers (Fig. 1a) intercalated by layers of S and In atoms, forming the alternating layers -In-[S-(Ni$_3$-In)-S]- (Fig. 1c). Owing to the weak van der Waals-like bonding between the S and In layers, cleaving predominantly produces two well-defined terminations: Indium (In)-terminated surface and Sulfur (S)-terminated surface (Fig. 1d). We distinguish these regions through spatially resolved micro-ARPES, which reveals a pronounced work-function difference of $\sim$0.61\,eV (see Supplementary Fig. 2 and Supplementary Fig. 3 for details), supplemented by scanning tunneling microscopy/spectroscopy (STM/S) measurements that further verifies the termination assignment. These combined probes enable an unambiguous identification of the surface terminations and allow us to resolve their distinct surface-dependent electronic structures. Notably, similar surface-dependent electronic features have also been reported in the ferromagnetic Weyl semimetal Co$_3$Sn$_2$S$_2$\cite{Morali2019,Huang2023,Mazzola2023,ekahana2025}, another shandite compound, where differences in electronic structure are primarily manifested in the different connection fashions for Fermi-arc surface states\cite{Morali2019,Mazzola2023}.

The three-dimensional (3D) bulk Brillouin zone and its projected (001) surface Brillouin zone for primitive cell and conventional cell of Ni$_3$In$_2$S$_2$ are illustrated in Fig. 1e,f with the high-symmetry points marked by red spots. Figure 1g (left panel) shows the calculated $k_z$-integrated bulk bands with SOC along the high-symmetry path $\bar{M}$--$\bar{\Gamma}$--$\bar{K}$--$\bar{M}$. By projecting the bands onto the Ni-3d orbitals, the flat band---a hallmark of kagome electronic structures---becomes clearly discernible. Unlike the ideally flat band (Fig. 1b) arising from a perfect kagome lattice, the kagome lattice-derived flat band in Ni$_3$In$_2$S$_2$ becomes dispersionless only within a limited momentum range around the $\bar{\Gamma}$ point. The density of states (DOS) calculation for bulk states in Ni$_3$In$_2$S$_2$ (right panel of Fig. 1g) reveals a prominent DOS peak at the flat-band energy, indicating that this feature originates primarily from Ni-3d orbitals. Moreover, the calculated real-space charge-density distribution of the flat-band at the $\bar{\Gamma}$ point (Supplementary Fig. 4a) confirms its localization around the Ni-based kagome network, providing further evidence that this flat band arises from destructive quantum interference of Bloch wavefunctions governed by the kagome lattice symmetry.

~\\
 \noindent {\bf Topological surface flat band and kagome lattice-derived flat band.} Besides the pronounced work-function difference between the two surface terminations, the S- and In-terminated surfaces also exhibit markedly different band structures, each displaying its own electronic features. High-resolution micro-ARPES measurements, supported by density functional theory (DFT) calculations, allow a direct comparison of the full electronic dispersions along the $\bar{M}$--$\bar{\Gamma}$--$\bar{K}$--$\bar{M}$ high-symmetry path for both surfaces (Fig. 2). The measured dispersions and their second-derivative plots (Fig. 2a,b,d,e) show excellent agreement with the calculated band structures, including surface-state contributions (Fig. 2c,f), establishing a solid foundation for identifying termination-specific low-energy features.

 On the S-terminated surface, we uncover a nearly nondispersive band appearing near the Fermi level along the $\bar{\Gamma}$--$\bar{M}$ direction (highlighted by the black arrows in Fig. 2b,g). This feature is faithfully reproduced in our surface-state electronic structure calculations (black arrow in Fig. 2c). Comparing the experimental and theoretical results, we find that this surface flat band lies within the bulk gap formed by the valence and conduction bands and extends almost entirely along the $\bar{\Gamma}$--$\bar{M}$ direction. In the actual material, when considering SOC, Ni$_3$In$_2$S$_2$ behaves as a weak topological insulator\cite{Zhang2022}. By referring to the topological invariants calculations and analyses ($z_{2,1}z_{2,2}z_{2,3}z_{2,4}$ = (1112)) by T. T. Zhang et al.\cite{Zhang2022}, this surface flat band can be assigned to a nontrivial surface state, which is topologically
 protected by bulk weak topological insulator via the bulk-boundary correspondence. Based on this, we introduce the concept of a Topological Surface Flat Band (TSFB) to describe it. 
 
 The TSFB feature can be quantitatively confirmed by comparing the integrated energy distribution curves (EDCs) along the $\bar{\Gamma}$--$\bar{M}$ high-symmetry direction for the S- and In-terminated surfaces (Fig. 4c). A well-defined TSFB peak is observed only on the S-terminated surface at a binding energy of $\sim$40\,meV, in close proximity to the Fermi level. Additionally, the TSFB exhibits an extremely narrow bandwidth of $\sim$30\,meV, obtained from the interval of peaks in EDCs at different momentum positions along the $\bar{\Gamma}$--$\bar{M}$ direction (see Supplementary Fig. 5d for details). Consistently, both our STM differential conductance spectrum (Fig. 4f) and DFT-calculated surface local density of states (LDOS) (Fig. 4i) reveal a pronounced DOS peak near the 40\,meV binding energy on the S-terminated surface, in excellent agreement with our ARPES observations. This suggests that the extremely narrow TSFB does indeed induce a sharp peak in the DOS near the Fermi level, significantly enhancing the electron-electron correlation effects. As illustrated in Fig. 2h, in contrast to the Dirac surface state observed in the seminal topological insulator Bi$_2$Se$_3$, which exhibits an extremely high electron velocity, the TSFB in Ni$_3$In$_2$S$_2$ is characterized by strongly quenched electron kinetic energy, leaving electron correlations dominant.

In addition to the TSFB, both terminations exhibit a prominent flat-band feature at a binding energy of $\sim$100\,meV near the $\bar{\Gamma}$ point (red arrows in Fig. 2b,e,g), further resolved in comprehensive high-resolution laser-ARPES measurements (Supplementary Fig. 6). By comparing with the calculated bulk band structures (Fig. 1g and Supplementary Fig. 5g), we identify that this flat band originates from geometric destructive interference of Bloch wavefunctions within the kagome layers in Ni$_3$In$_2$S$_2$ (schematically shown in Fig. 2i), and it possesses an ultranarrow bandwidth of $\sim$5\,meV, as analyzed from the EDC stacks (Supplementary Fig. 5h). To the best of our knowledge, this constitutes the first direct spectroscopic observation of a kagome lattice-derived flat band in the “322” kagome family. 

Orbital-projected calculations (Supplementary Fig. 4b), supported by polarization-dependent laser-ARPES measurements (Supplementary Fig. 6), reveal that this flat band is mainly composed of the out-of-plane Ni-d$_{z^2}$ orbitals, with additional hybridization from the in-plane Ni-d$_{x^2-y^2/xy}$ and Ni-d$_{xz/yz}$ orbital components. This out-of-plane orbital character contrasts sharply with the flat bands reported in kagome metals such as CoSn\cite{Kang2020b} and CsTi$_3$Bi$_5$\cite{Yang2023}, which are dominated by in-plane orbital contributions. Furthermore, unlike the ideal kagome metal, in which the flat-band spectral weight spreads uniformly across the two-dimensional (2D) Brillouin zone\cite{Kang2020a}, the flat band in Ni$_3$In$_2$S$_2$ only emerges in a restricted region around $\bar{\Gamma}$. Our laser-ARPES intensity maps reveal that its spectral weight forms a warped triangular patch exhibiting clear threefold symmetry (highlighted by the yellow-shaded region in the inset of Fig. 2g and Supplementary Fig. 7 for details). This localized, $C_3$-symmetric distribution originates from the realistic lattice environment and symmetry of Ni$_3$In$_2$S$_2$, where the unusual three-layer kagome stacking per unit cell builds an interacting kagome system, while interlayer coupling, SOC, and orbital hybridization collectively drive substantial deviations from the ideal kagome material.

~\\
 \noindent {\bf 2D Dirac-node arc state on the In-terminated surface.} Unlike the S-terminated surface, where a nontrivial TSFB appears, the In-terminated surface does not exhibit such a feature. However, we notice that the Dirac dispersion near the $\bar{M}$ point along the $\bar{K}$--$\bar{M}$--$\bar{K}$ direction on the In-terminated surface (Fig. 2e) shows a clear gapless character. The similar Dirac dispersion is also seen to exist on the S-terminated surface, but is apparently gapped (Fig. 2b). This key difference is further confirmed by comparing the EDCs extracted at the $\bar{M}$ point for both surfaces, as shown in Supplementary Fig. 8: while the In-terminated surface exhibits a gapless Dirac dispersion, the Dirac point on the S-terminated surface is gapped by $\sim$76\,meV. To verify whether the Dirac bands observed on the two terminations are surface-originated, we conducted photon-energy--dependent ARPES measurements along the $\bar{\Gamma}$--$\bar{K}$--$\bar{M}$ high-symmetry path for both the In- and S-terminated surfaces, thereby varying the out-of-plane momentum $k_z$. The results (see Supplementary Fig. 9) indicate that the Dirac states at the $\bar{M}$ point on both surfaces exhibit no observable $k_z$ dispersion, confirming their 2D or surface-state nature.

Remarkably, our further measurements reveal that the electronic states near the $\bar{M}$ point on the In-terminated surface form a rare 2D Dirac-node arc state. We employed synchrotron-based micro-ARPES ($h\nu$ = 34\,eV) to investigate this unusual state. Figures 3a,b present constant-energy contour (CEC) maps at different binding energies $E_B$. Along the $\bar{\Gamma}$--$\bar{M}$ direction, six parallel line-pairs (outlined by green dashed lines) emerge. As the binding energy increases, these features gradually shrink, merge and disappear. At the Dirac-point energy, the line-pair merges into a single line. To trace the evolution of the Dirac states near the $\bar{M}$ point, we extract band dispersions along the cuts as marked by dashed lines on the Fermi surface (Fig. 3e). The obtained band dispersions, along the paths parallel to the $\bar{K}$--$\bar{M}$--$\bar{K}$ direction, illustrate how the surface Dirac states evolve from cuts 1 to 6 (Fig. 3d). Along the cut 4 which passes exactly through the $\bar{M}$ point, it exhibits a sharp, gapless, linear Dirac dispersion, crossing at $E_B$ = 0.18\,eV. Momentum distribution curves (MDCs) fitting (Supplementary Fig. 10) yields the typical full width at half maximum (FWHM) of only $\sim$0.04\,$\mathrm{\AA^{-1}}$. The extracted Fermi velocity ($V_F$) is as high as 3.974\,$\pm$\,0.006\,$\times$\,10$^{5}$\,m/s, comparable to the corresponding values reported in well-known Dirac-electron materials such as graphene\cite{siegel2011}, Bi$_2$Se$_3$\cite{Xia2009}, and MnBi$_2$Te$_4$\cite{Hao2019}. On both sides of the $\bar{M}$ point, the Dirac point energy slowly shifts downward, and a finite gap opens in cuts 1, 2 and 6. As shown in Fig. 3g, we compare the EDCs extracted at the Dirac points from cuts 1--6, visualizing the evolution from gapped to gapless and back to gapped Dirac states. The detailed $k_y$ dependence of the Dirac point energy and gap size along the $\bar{\Gamma}$--$\bar{M}$ direction is plotted in Fig. 3h. As $k_y$ approaches the $\bar{M}$ point (at $k_y$ = 0.6755\,$\mathrm{\AA^{-1}}$), the gap closes completely at around $k_y$ = 0.58\,$\mathrm{\AA^{-1}}$, resulting in a well-defined gapless state. We therefore conclude that the gapless surface Dirac nodes around the time-reversal invariant momentum (TRIM) point $\bar{M}$ are connected together in momentum space, forming a finite-length arc---the 2D Dirac-node arc state. To further corroborate this result, we performed higher-resolution, higher-statistics and laser-based ARPES measurements, which clearly resolve the transition from gapped to gapless Dirac states (Fig. 3i and Supplementary Fig. 11). Additionally, we investigated the circular dichroism (CD) of this 2D Dirac-node arc state using left- and right-circularly polarized laser light. The intensity contrast across the mirror plane exhibits a reversal both above and below the Dirac point (Supplementary Fig. 12), similar to the CD response observed in the topological Dirac surface states of Bi$_2$Se$_3$\cite{Wang2011,Park2012}. This implies possible orbital angular momentum (OAM) texture, spin-momentum locking, or other nontrivial spin-orbital entanglement. 

To better illustrate this novel quantum state, we plotted the distribution of Dirac nodes in momentum space (Fig. 3f). A 3D schematic of the Dirac-node arc structure is also shown in Fig. 3c, where the gapless Dirac points extend along an open one-dimensional line in momentum space. Together with the photon-energy--dependent measurements of the Dirac dispersion at the $\bar{M}$ point (Supplementary Fig. 9), our data confirm that this is a truly 2D Dirac-node arc state---a remarkably rare type of surface electronic state, previously reported only in PtSn$_4$\cite{Wu2016}, HfSiS\cite{Takane2016}, and Hf$_2$Te$_2$P\cite{Hosen2018}. This exotic state likely underlies the unusual transport behaviour of Ni$_3$In$_2$S$_2$, including its record-high carrier mobility among all known kagome materials as well as large and anisotropic magnetoresistance.

~\\
\noindent {\bf Atomic-scale visualization and QPI signature in Ni$_3$In$_2$S$_2$.} 
To directly confirm our cleavage terminations assignment in ARPES measurements, we performed atomic-scale STM imaging on cleaved Ni$_3$In$_2$S$_2$ crystals. The STM topography clearly reveals two distinct surface types, one with vacancies and one with adatoms. When the two surfaces meet at a step edge, the upper surface is In-terminated surface with In vacancies and the lower one is S-terminated surface with In adatoms (Supplementary Fig. 13), which is dictated by crystal configuration. We argue the atomic step-edge geometry imaging is a generic methodology for surface determination, and is successfully applied in various kagome material families\cite{Liu2024Perspective,liu2025atomic}.

We further performed quasiparticle interference (QPI) measurements on the two distinct surfaces and compared them with the corresponding Fermi surfaces obtained from ARPES and calculations. Both the QPI pattern and Fermi surface from S surface exhibit parallel line-pairs along the $\bar{\Gamma}$--$\bar{M}$ direction and a short convex arc-line pair near the center between the $\bar{\Gamma}$ and $\bar{K}$ points, while the In surface shows a long concave arc, as indicated by the red marks in Fig. 4a,b,d,e. To further interpret these results, we calculated the Fermi surface maps for both surface terminations (Fig. 4g,h). The Fermi surface geometries are markedly different between the two terminations, yet both agree well with the experimental observations (see also Supplementary Fig. 14 for side-by-side comparisons). 
The good agreement between theory and experiment further confirms the surface origin and momentum-resolved characteristics of the band structures revealed by ARPES and STM on Ni$_3$In$_2$S$_2$.

~\\
\noindent{\bf\large{Discussion}}\\
Flat bands are relatively uncommon, and even rarer are the surface flat bands protected by nontrivial topology. Earlier efforts mainly focused on nodal-line semimetals, where nearly nondispersive “drumhead” surface states (DSs) are predicted to appear within the projected regions of the bulk nodal lines\cite{Weng2015}. In real materials, however, SOC typically gaps the nodal lines into discrete nodal points, making such surface states difficult to stabilize and even more difficult to be detected due to their limited momentum-space extension. Although previous studies have reported experimental signatures of drumhead states\cite{Bian2016,Belopolski2019,LesliePRX}, a genuine observation of topological surface flat bands with large momentum-space extension and minimal bandwidth remains elusive. Only recently have nearly ideal drumhead surface flat bands been observed in rhombohedral graphite\cite{Zhou2024,Xiao2025}, possibly accompanied by strong-correlation effects. However, due to the lack of robust topological protection, even minor surface perturbations can easily destroy these drumhead states\cite{Fang2016}. Therefore, identifying the genuinely and topologically protected surface flat bands constitutes a new frontier in flat-band physics. 

In this context, our study demonstrates a robust TSFB on the S-terminated surface of bulk Ni$_3$In$_2$S$_2$, located at $\sim$40\,meV below the $E_F$, revealed through high-resolution ARPES, STM, and DFT calculations. This TSFB originates from the topologically protected surface states inherent to a weak topological insulator phase (the nonzero topological invariants of $z_{2,1}z_{2,2}z_{2,3}z_{2,4}$ = (1112)). Unlike the strong topological insulators such as Bi$_2$Se$_3$, which possess sizable bulk gaps between conduction and valence bands throughout momentum space, the conduction and valence bands in the topological kagome metal Ni$_3$In$_2$S$_2$ intersect each other near the $\bar{\Gamma}$ point. As a result, the originally highly dispersive Dirac surface states become dramatically compressed, generating a TSFB within a bulk gap. This represents a fundamentally new mechanism for topological surface flat-band formation, significantly broadening the platform for flat-band research. 

The topological protection, combined with the large DOS near $E_F$ arising from the flat dispersion, positions the TSFB as a promising platform for exploring interaction-driven spontaneous symmetry breaking and topological phase transitions. External tuning parameters, such as doping and strain, could provide controlled tuning of the TSFB and thereby facilitate the exploration of intricate couplings between flat-band correlations and nontrivial topological physics.

In addition, Ni$_3$In$_2$S$_2$, as a newly discovered “322” kagome material isostructural to Co$_3$Sn$_2$S$_2$, provides an important reference system for comprehensively understanding kagome-lattice physics. We have directly observed an exceptionally narrow kagome lattice-derived flat band located only $\sim$100\,meV below the Fermi level---substantially closer to $E_F$ than the flat bands previously identified in FeSn, CoSn, and CsTi$_3$Bi$_5$ (typically 200--300\,meV below $E_F$)\cite{Kang2020a,Kang2020b,Yang2023}. Such a closing in energy, combined with its ultranarrow bandwidth of $\sim$5\,meV, provides an ideal opportunity for exploring correlation-driven topological phases emerging from the partially filled kagome flat bands. Despite its kagome origin, this flat band exhibits two striking deviations from the ideal kagome limit: (i) it is dominated by out-of-plane d$_{z^2}$-orbital character, in sharp contrast to the flat bands in CoSn\cite{Kang2020b} and CsTi$_3$Bi$_5$\cite{Yang2023} that originate from in-plane orbital components, and (ii) its spectral weight is confined to a finite, $C_3$-symmetric triangular region around $\bar{\Gamma}$ rather than being uniformly distributed across the full 2D Brillouin zone, as observed in the ideal kagome metal FeSn\cite{Kang2020a}. These characteristics reflect the strong interlayer coupling, orbital hybridization, and realistic stacking complexity inherent to the “322” kagome lattice, explaining why clear ARPES evidence of kagome lattice-derived electronic states has long remained elusive in the family of “322” kagome materials. In this sense, Ni$_3$In$_2$S$_2$ represents a strongly interacting counterpart to the ideal kagome system, substantially enriching the broader landscape of kagome flat-band physics.

Although the kagome flat band in the “322” kagome family has been challenging to be detected, multiple experimental signatures have suggested its existence and influence on the electronic properties. In Co$_3$Sn$_2$S$_2$, resonant inelastic x-ray scattering (RIXS) measurements\cite{Nag2022} recently identified a nearly nondispersive Stoner excitation at $\sim$0.38\,eV with a very narrow bandwidth of $\sim$0.05\,eV, attributed to the underlying kagome flat band, and this excitation persists well above the Curie temperature, revealing a strong coupling between flat-band--driven correlation effects and spin polarization, and underscoring the pivotal role of flat bands in correlated spin dynamics. STM measurements have further uncovered unconventional negative orbital magnetism linked to the kagome lattice-derived flat band in Co$_3$Sn$_2$S$_2$\cite{Yin2019}. In the structurally related kagome compound Ni$_3$In (which shares the same Ni-based kagome lattice as Ni$_3$In$_2$S$_2$), the transport and STM/STS experiments indicate that the partial filling of a Ni-3d kagome flat band drives anomalous strange-metal behaviour\cite{Ye2024,Souza2025}. These observations collectively highlight the central role of kagome lattice-derived flat bands in mediating correlated and magnetic responses across related kagome systems. In this context, our ARPES measurements in Ni$_3$In$_2$S$_2$ provide the first direct spectroscopic observation of a momentum-space destructive-interference--derived flat band in the “322” kagome family, filling a long-standing experimental gap and establishing a new platform for exploring flat-band--induced magnetism, correlation effects, and topological phenomena in kagome materials.

Given that both the TSFB and the kagome lattice-derived flat band in Ni$_3$In$_2$S$_2$ lie in close proximity to the Fermi level ($\sim$40\,meV and $\sim$100\,meV below $E_F$, respectively), tuning the flat bands closer to the Fermi level through element substitution or electrostatic gating could be an effective step toward realizing flat-band--driven correlated phases. For example, hole doping in Ni$_3$In$_2$S$_2$ can be introduced by substituting Ni with Fe or Co (e.g., Ni$_{3-x}$Fe$_x$In$_2$S$_2$ or Ni$_{3-x}$Co$_x$In$_2$S$_2$), which is expected to significantly modify the energy position of the flat bands and potentially induce new emergent behaviour---an approach already demonstrated in Co$_3$Sn$_2$S$_2$\cite{Mangelis2017,Lohani2023}. Additionally, recent advances in gating technology have enabled the precise control of the electronic density in micro-fabricated Co$_3$Sn$_2$S$_2$ devices, with the Fermi level shifting by up to 200\,meV\cite{Matsuoka2025}. These developments suggest that applying similar strategies to Ni$_3$In$_2$S$_2$ could enable direct and finely tunable control of its flat-band states, thereby creating a powerful platform for engineering correlation-driven and topological phases in kagome materials.

Of particular interest is the discovery of a rare 2D Dirac-node arc state on the In-terminated surface, which unveils a previously unexplored landscape of Dirac physics in kagome materials. This state appears as a finite-length arc near the $\bar{M}$ point on the Brillouin-zone boundary, formed by gapless Dirac nodes connected along the $\bar{\Gamma}$--$\bar{M}$ direction. Although the theoretical calculations of surface states on the In-terminated surface do not fully capture this node arc state, a comparison with the bulk band projections reveals a Dirac dispersion near the $\bar{M}$ point (Supplementary Fig. 10c), with the Dirac-point energy and Fermi velocity being in reasonable agreement with those observed experimentally for the Dirac surface state. Therefore, we attribute this 2D Dirac-node arc state to a unique surface resonance state (SRS) on the In-terminated surface, which inherits the bulk band dispersion but exhibits enhanced spectral weight at the surface\cite{Zangwill1988}. Similar Dirac dispersions formed by SRS have been reported in materials such as PdTe$_2$\cite{bahramy2018ubiquitous} and TaPdTe$_5$\cite{lu2022topologically}. In Ni$_3$In$_2$S$_2$, however, such a SRS displays an unconventional strong CD, suggesting a nontrivial OAM texture or spin-orbit entanglement characteristics in this kagome-based compound. Importantly, the 2D Dirac node-arc provides a linearly dispersive channel with an ultra-high Fermi velocity parallel to $\bar{K}$--$\bar{M}$--$\bar{K}$ direction. Owing to the coupling between the SRS and bulk states, this massless Dirac dispersion can penetrate several atomic layers beneath the surface and form a surface-assisted bulk conduction channel that potentially supports high-mobility charge transport \cite{Fang2023,Zhang2024,KimPRB}. Both the SRS and bulk-coupled Dirac carriers are thus expected to contribute to the observed anisotropic magnetoresistance\cite{Zhang2022}. The small effective mass and likely long relaxation time of these Dirac carriers, as supported by the sharp Dirac dispersion shown in Supplementary Fig. 10, could underlie the extremely large transverse magnetoresistance dominated by Lorentz deflection of high-mobility carriers. 
The 2D Dirac-node arc state, arising from the unique surface resonance state, not only provides a new route for fundamental exploration of bulk-surface coupled Dirac physics but also opens exciting opportunities for expanding the application prospects of high-performance Dirac electronics in kagome materials.


In conclusion, we have systematically mapped the electronic landscape of the newly discovered “322” kagome metal Ni$_3$In$_2$S$_2$. On the S-terminated surface, we reveal a robust TSFB located only $\sim$40\,meV below $E_F$, originating from nontrivial weak topological insulator surface states, representing a novel mechanism for topological flat-band formation. Importantly, our high-resolution ARPES measurements also provide the first direct spectroscopic evidence of a kagome lattice-derived flat band in the bulk of the M$_3$A$_2$B$_2$ kagome compound, located at a binding energy of $\sim$100\,meV with an ultranarrow bandwidth of $\sim$5\,meV, thereby filling a long-standing gap in the direct experimental confirmation of kagome flat bands in the “322” kagome family. Furthermore, on the In-terminated surface we uncover an exceptionally rare 2D Dirac-node arc state, characterized by sharp linear dispersion, exceptionally high Fermi velocity, and pronounced CD. The unprecedented coexistence of these nontrivial electronic states thus reveals a deep intertwining between flat-band correlations and topological effects in Ni$_3$In$_2$S$_2$. These findings not only broaden the fundamental understanding of correlated and topological phenomena in kagome systems but also establish an experimental foundation for designing next-generation topological quantum materials hosting flat-band physics. 

~\\
\noindent{\bf\large Methods}

\noindent{\bf Growth of single crystals.} Ni$_3$In$_2$S$_2$ single crystals were successfully synthesized using self-flux method\cite{Kumar2023}. High-purity elements of Nickel lumps (3\,mol, 0.75\,g), Indium balls(2\,mol, 0.979\,g) and Sulphur pieces(2\,mol, 0.283\,g) were mixed and placed in an alumina crucible. The crucible was then sealed in a quartz tube under high-vacuum conditions. The tube was heated to 400\,$^\circ$C within 10 hours and held for 5 hours to prevent sulfur volatilization loss due to excessive heating rate. Subsequently, the tube was heated to 1000\,$^\circ$C and held for 30 hours. Then it was cooled to 800\,$^\circ$C at a rate of 1.5\,$^\circ$C/hour. The muffle furnace was then turned off and allowed to cool naturally to room temperature. Thin flakes of Ni$_3$In$_2$S$_2$ single crystals were mechanically cleaved from the obtained metal ingot.\\ 


\noindent{\bf ARPES measurements.} Synchrotron-based ARPES measurements were performed with micro-focused VUV light at the 03U beamline of Shanghai Synchrotron Radiation Facility (SSRF)\cite{Yang2021} with a hemispherical electron energy analyzer DA30L (Scienta-Omicron). The energy and momentum resolutions were set to better than 20\,meV and 0.02\,$\mathrm{\AA^{-1}}$. The light spot size was smaller than $20\,\mu\mathrm{m}$. High-resolution ARPES measurements were also performed using a lab-based ARPES system equipped with the micro-focused 6.994\,eV vacuum-ultra-violet (VUV) laser and a hemispherical electron energy analyzer DA30L (Scienta-Omicron)\cite{Liu2008,Zhou2018}. The laser spot was focused to around $10\,\mu\mathrm{m}$ on the sample in order to distinguish different surface terminations. The energy resolution was set to better than 1.5\,meV in most cases, and the angular resolution was $\sim$0.1$^\circ$ corresponding to the momentum resolution of $\sim$0.001\,$\mathrm{\AA^{-1}}$. The preliminary ARPES measurements were carried out on the BL13U beamline of the National Synchrotron Radiation Laboratory (NSRL) and the BL-9B of Hiroshima Synchrotron Radiation Center (HiSOR). All the samples were cleaved in situ at a low temperature of 15\,K and measured in ultra-high vacuum with a base pressure better than 6$\times$10$^{-11}$\,mbar. The Fermi level is referenced by measuring on a clean polycrystalline gold that is electrically connected to the sample.\\

\noindent{\bf STM measurements.} Single crystal Ni$_3$In$_2$S$_2$ was mechanically cleaved in situ at 77\,K under ultrahigh-vacuum conditions, and immediately transferred into the scanning tunneling microscopy head maintained at the base temperature of liquid He$^4$ (4.2\,K). We extensively scanned each crystal to identify large, clean S- and In-terminated surfaces. Typically, it takes one week or more to find micrometer-size clean surfaces with $<$ 1\% In adatoms or In vacancies, where we can obtain sharp QPI signals. Tunneling conductance spectra were acquired at $V$ = -100\,mV and $I$ = 0.5\,nA using a standard lock-in amplifier technique, with a modulation voltage of 5\,mV and a lock-in frequency of 973.231\,Hz. QPI measurements are taken with the junction set-up: $V$ = 20\,mV, $I$ = 0.3\,nA, and a modulation voltage of 3\,mV.\\

\noindent{\bf Electronic structure calculations.} The first-principles calculations are performed based on DFT\cite{DFT1} within the Perdew-Burke-Ernzerhof (PBE) exchange-correlation\cite{DFT2} using the Vienna {\it ab initio} simulation package (VASP)\cite{DFT3}. The plane-wave cutoff energy is set to be 400\,eV with a 5$\times$5$\times$2 k-mesh in the Brillouin zone for the self-consistent calculations. All calculations in this work, unless otherwise specified, include SOC. Based on the results of the DFT calculations, we construct the maximally localized Wannier functions (MLWFs)\cite{WANNIER1,WANNIER2} using the WANNIER90 package\cite{WANNIER3}. During the construction of the MLWFs, we select the following orbitals: Ni-d, S-p, In-s and In-p for the Ni$_3$In$_2$S$_2$. After that we utilize the WannierTools package\cite{WANNIER4} to calculate the Fermi surface and surface state of Ni$_3$In$_2$S$_2$. The lattice constants used in our calculation are a = b = 5.37\,$\mathrm{\AA}$ and c = 13.56\,$\mathrm{\AA}$\cite{Zhang2022}.

~\\
\noindent {\bf\large Data availability}\\
All data are processed by using Igor Pro 8.02 software. All data needed to evaluate the conclusions in the paper are available within the article and its Supplementary Information files. All raw data generated during the current study are available from the corresponding author upon reasonable request.

~\\
\noindent {\bf\large Code availability}\\
The codes used for the DFT calculations in this study are available from
the corresponding authors upon request.

~\\
\noindent {\bf\large References}

\bibliographystyle{naturemag}
\bibliography{NISRef}


\vspace{3mm}

\noindent {\bf\large Acknowledgements}\\
We thank Prof. Hongming Weng for fruitful discussions and Prof. Zhaoxiang Wang and Mengyan Cao for their assistance with the Energy-dispersive X-ray spectroscopy (EDX) measurements. This work is supported by the National Key Research and Development Program of China (Grants No. 2023YFA1406103 and 2022YFA1403901 by G.D.L., 2024YFA1409100 and 2021YFA1400400 by H.J.Z., 2024YFA1408400 by Y.G.S. and 2022YFA1604200 by L. Z.), the National Natural Science Foundation of China (Grants No. 12488201 by X.J.Z., 12374066 by L.Z., 12534007 and 92365203 by H.J.Z.), the Natural Science Foundation of Jiangsu Province (Grants No. BK20233001, No. BK20253012 and No. BK20243011 by H.J.Z.), the e-Science Center of Collaborative Innovation Center of Advanced Microstructures, and the Synergetic Extreme Condition User Facility (SECUF, https://cstr.cn/31123.02.SECUF). We thank the Shanghai Synchrotron Radiation Facility of BL03U(https://cstr.cn/31124.02.SSRF.BL03U) for the assistance on ARPES measurements. We thank the staff members of the ARPES System(https://cstr.cn/31131.02.HLS.ARPES) at the National Synchrotron Radiation Laboratory in Hefei (https://cstr.cn/31131.02.HLS), for providing technical support and assistance in data collection and analysis. Synchrotron ARPES measurements at HiSOR were performed under the approval of the Program Advisory Committee (Proposal Numbers: 22BG045 and 23AG023).

\vspace{3mm}

\noindent {\bf\large Author Contributions}\\
 B.L., Y.C.L., J.P. and H.B.D. contribute equally to this work. G.D.L., X.J.Z., B.L., and H.J.Z. proposed and designed the research. J.P. and Y.G.S. contributed to single crystal growth and characterizations. L.H.Y. assisted in the XRD characterization. Y.C.L. and H.J.Z. contributed to the DFT band calculations. T.T.Z. and Y.T.Z. contributed to the discussion of the DFT and STM results. H.B.D. conducted the STM experiments in consultation with J.-X.Y.; B.L. carried out the ARPES experiment. B.L., T.M.M., W.P.Z., N.C., M.K.X., H.C., X.L.R., C.H.Y., Y.J.S., Y.W.C., F.F.Z., S.J.Z., L.Z., Z.Y.X., G.D.L. and X.J.Z. contributed to the development and maintenance of Laser-ARPES systems. J.Y.L., Z.C.J., Z.T.L., D.W.S. and M.Y. contributed to the experimental assistance in SSRF. Z.F.L., H.E.Z., Y.L.L., T.R.L., S.T.C. and Z.S. contributed to the experimental assistance in NSRL. K.M., T.O. and K.S. contributed to the experimental assistance in HiSOR. B.L., G.D.L., Y.C.L. and H.J.Z. analyzed the data. B.L., G.D.L. and H.B.D. wrote the paper. All authors participated in the discussion and comment on the paper.

\vspace{3mm}

\noindent{\bf\large Competing interests}\\
The authors declare no competing interests.

\newpage

\begin{figure*}[tbp]
\begin{center}
\includegraphics[width=1.0\columnwidth,angle=0]{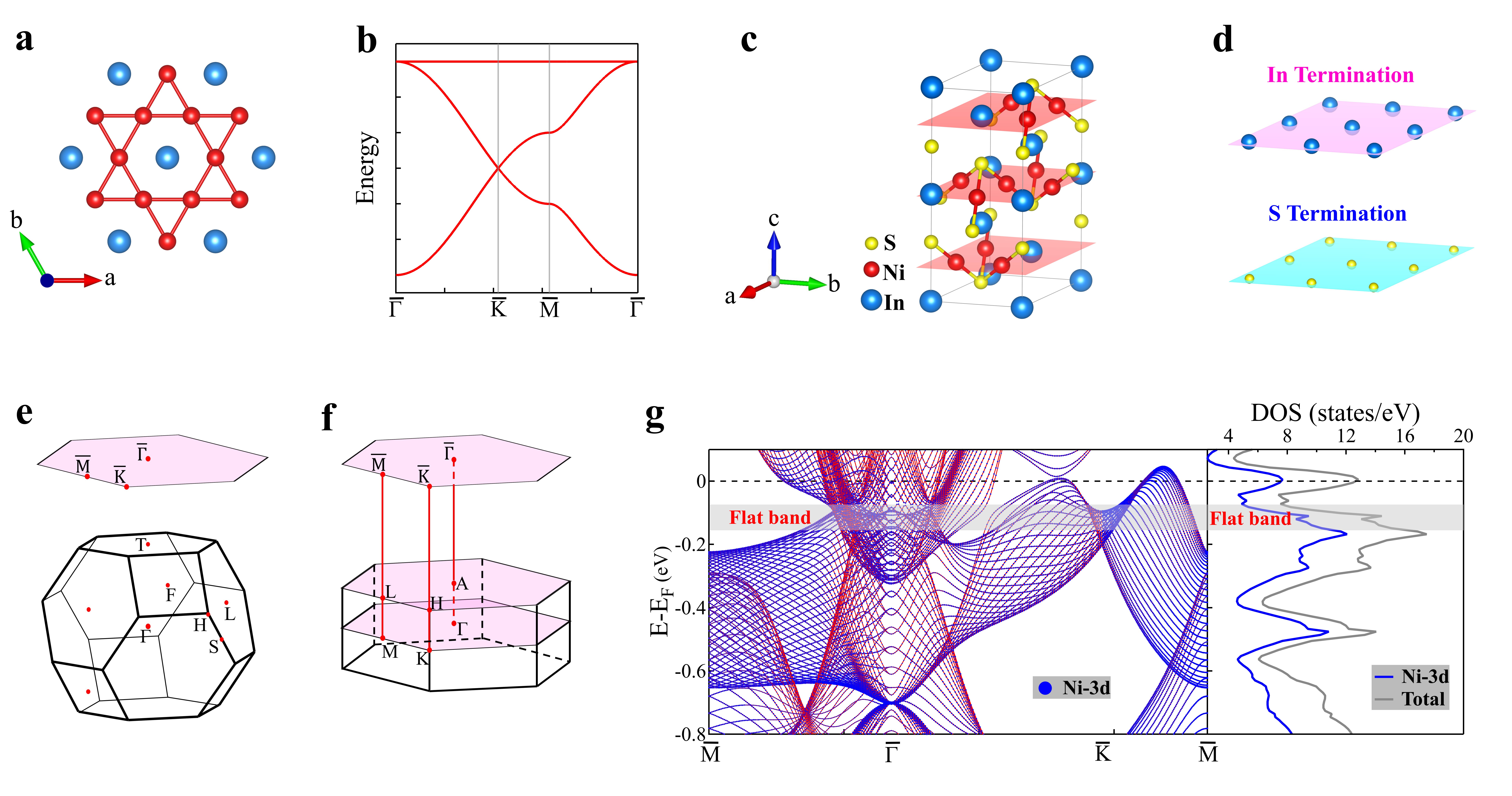}
\end{center}
\caption{\textbf{Lattice structure and calculated bulk electronic structure of Ni$_3$In$_2$S$_2$.} \textbf{a} Top view of the kagome lattice composed of nickel atoms. \textbf{b} The ideal kagome lattice band dispersion considering only the nearest-neighbor hopping.
\textbf{c} A unit cell of Ni$_3$In$_2$S$_2$ with three layers of kagome planes highlighted by the red squares.
\textbf{d} Two possible cleaving surfaces corresponding to the In- and S-terminated layers, whose atoms are arranged in a triangular fashion.
\textbf{e} Schematic plot of the bulk Brillouin zone of the primitive cell and its projection forming the (001) surface Brillouin zone.
\textbf{f} The conventional cell Brillouin zone and projected (001) surface Brillouin zone.
\textbf{g} DFT-calculated bulk bands of Ni-3d orbital integrated over all $k_z$ (left panel) and the DOS calculation for bulk states in Ni$_3$In$_2$S$_2$ (right panel). The grey shading bar in (\textbf{g}) highlights the flat band derived from the Ni-based kagome network.
}
\end{figure*}

\begin{figure*}[tbp]
\begin{center}
\includegraphics[width=1.0\columnwidth,angle=0]{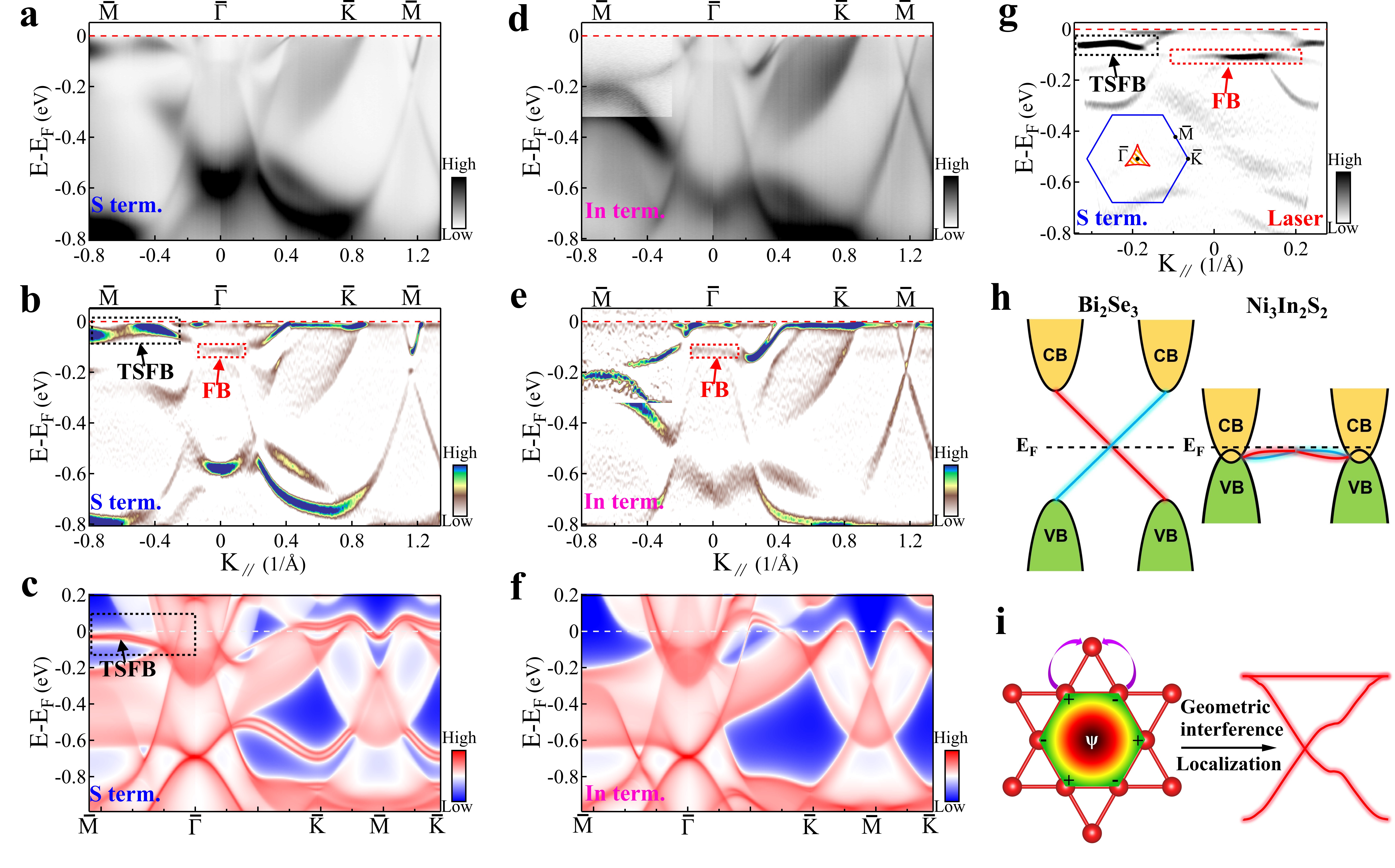}
\end{center}
\caption{\textbf{Signature of TSFB and kagome lattice-derived flat band in Ni$_3$In$_2$S$_2$.} \textbf{a--c} For the S-terminated surface (Top to bottom) with ARPES spectra along the $\bar{M}$--$\bar{\Gamma}$--$\bar{K}$--$\bar{M}$ high symmetry directions (\textbf{a}), the corresponding second-derivative image with respect to energy (\textbf{b}), and the calculated band structure with surface states (\textbf{c}). The observed TSFB and kagome lattice-derived flat band are marked by black and red arrows, respectively. \textbf{d--f} Same as (\textbf{a--c}) but for the In-terminated surface. The original spectra for the two surfaces in panels (\textbf{a}) and (\textbf{d}) were both measured using linear horizontal (LH) polarized light with a photon energy of 34\,eV at 10\,K. \textbf{g} Second-derivative plot with respect to energy of high-resolution laser-based ARPES spectrum along the partial $\bar{\Gamma}$--$\bar{M}$ high symmetry line, obtained at 15\,K by using linear vertical (LV) polarized light with a photon energy of 6.994\,eV. TSFB and FB features were observed more clearly in the laser data (also indicated by black and red arrows respectively). The kagome flat band distribution in the 2D Brillouin zone is displayed in the inset of (\textbf{g}), which shows an interesting concave triangle. \textbf{h} Sketch of the TSFB in Ni$_3$In$_2$S$_2$, in contrast to the typical Dirac surface state in a seminal topological insulator Bi$_2$Se$_3$, which has an extremely high Fermi velocity. 
\textbf{i} The schematic diagram of the destructive interference of electron wave functions within the kagome lattice, and the resulting feature of flat band in the momentum space.
}
\end{figure*}

\begin{figure*}[tbp]
\begin{center}
\includegraphics[width=1.0\columnwidth,angle=0]{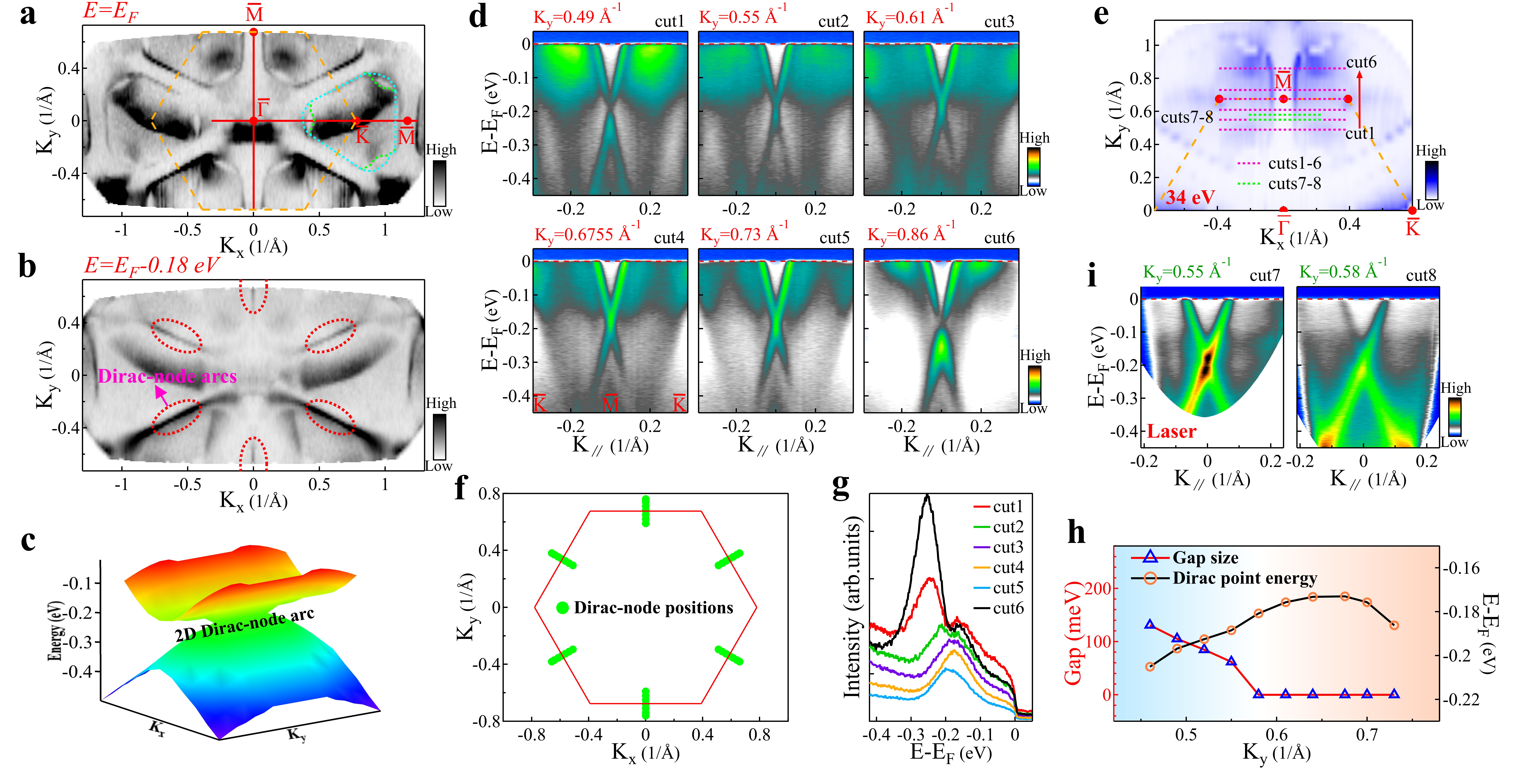}
\end{center}
\caption{\textbf{2D Dirac-node arc state on the In-terminated surface}. \textbf{a} Fermi surface and \textbf{b} Constant energy contour at the binding energy of 0.18\,eV, with Dirac nodes showing up around the $\bar{M}$ point. They are obtained by integrating the spectral intensity within 10\,meV with respect to the $E_F$ and $E_F$\,-\,0.18\,eV, respectively. 
\textbf{c} Schematic illustration of the 2D Dirac-node-arc structure as revealed by ARPES. 
\textbf{d} Band dispersions around the $\bar{M}$ point along cuts 1--6, whose locations are indicated by the pink dashed lines in panel (\textbf{e}).
\textbf{e} Zoomed-in view of the measured Fermi surface with partial surface Brillouin zone and high symmetry momentum points overlaid.
\textbf{f} Quantitative momentum locations shown as green bars of the observed Dirac nodes around the $\bar{M}$ point.
\textbf{g} EDCs stacked at the momentum positions of Dirac points in (\textbf{d}), illustrating gapless Dirac dispersions (cuts 3--5) and gapped ones (cuts 1, 2, and 6).
\textbf{h} Evolution of the gap size and Dirac point energy with the increasing $k_y$ close to the $\bar{M}$ point. The data in panels (\textbf{a--h}) were acquired at 10\,K by using synchrotron-based ARPES with a photon energy of 34\,eV.
\textbf{i} Band dispersions measured at 15\,K by using laser-based ARPES with a photon energy of 6.994\,eV. The cuts 7--8 parallel to the $\bar{K}$--$\bar{M}$--$\bar{K}$ direction are marked with green dotted lines in (\textbf{e}).
 }

\end{figure*}

\begin{figure*}[tbp]
\begin{center}
\includegraphics[width=1.0\columnwidth,angle=0]{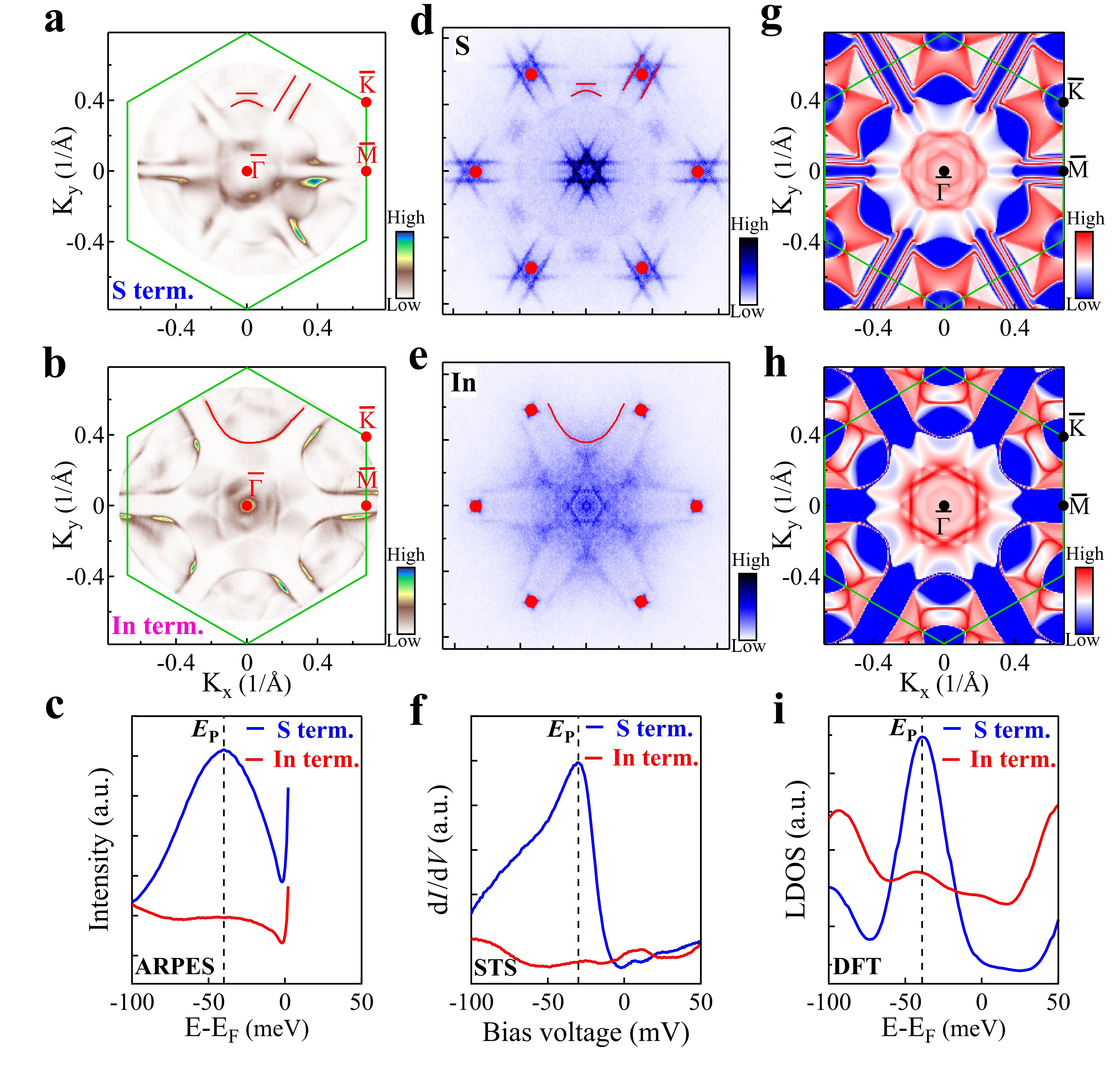}
\end{center}
\caption{\textbf{Fermi surface and QPI signature of the two types of cleavage surfaces in Ni$_3$In$_2$S$_2$}. 
\textbf{a, b} Experimental Fermi surfaces for the S- (\textbf{a}) and In- (\textbf{b}) terminated surfaces, respectively, measured at 15\,K by using laser-based ARPES (6.994\,eV). \textbf{c} EDCs obtained by integrating the ARPES data of the entire $\bar{\Gamma}$--$\bar{M}$ path shown in Fig. 2a,d. The data have been divided by the Fermi--Dirac distribution. 
\textbf{d, e} QPI patterns acquired at +20\,meV under zero magnetic field for the S- (\textbf{d}) and In- (\textbf{e}) terminated surfaces. Bragg points are marked by red dots.
\textbf{f} Differential conductance spectrum ($dI/dV$) taken on two terminations. 
\textbf{g, h} Calculated Fermi surface maps with surface states of S- (\textbf{g}) and In- (\textbf{h}) terminated surfaces.
\textbf{i} Calculated LDOS on two terminations. The similar arc-like features captured by ARPES and QPI are marked by red arcs in (\textbf{a}, \textbf{b}, \textbf{d}, \textbf{e}).
} 	
\end{figure*}

\end{document}